\documentclass[preprint,aps]{revtex4}

\usepackage{dcolumn}% Align table columns on decimal point
\usepackage{bm}% bold math

\begin{document}

%\preprint{}

\title{Unequal arm space-borne gravitational wave detectors}

\author{Shane L. Larson}
     \email{shane@srl.caltech.edu}
     \affiliation{Space Radiation Laboratory \\ California Institute of
Technology, Pasadena, CA 91125}

\author{Ronald W.\ Hellings}
     \email{hellings@physics.montana.edu}
\author{William A.\ Hiscock}
     \email{hiscock@montana.edu}
\affiliation{Department of Physics, Montana State University \\ Bozeman,
Montana 59717}

\date{\today}

\begin{abstract}
Unlike ground-based interferometric gravitational wave detectors,
large space-based systems will not be rigid structures.  When the
end-stations of the laser interferometer are freely flying
spacecraft, the armlengths will change due to variations in the
spacecraft positions along their orbital trajectories, so the
precise equality of the arms that is required in a laboratory
interferometer to cancel laser phase noise is not possible.
However, using a method discovered by Tinto and Armstrong, a
signal can be constructed in which laser phase noise exactly
cancels out, even in an unequal arm interferometer.  We examine
the case where the ratio of the armlengths is a variable
parameter, and compute the averaged gravitational wave transfer
function as a function of that parameter.  Example sensitivity
curve calculations are presented for the expected design
parameters of the proposed LISA interferometer, comparing it to a
similar instrument with one arm shortened by a factor of 100, showing
how the ratio of the armlengths will affect the overall sensitivity of the
instrument.
\end{abstract}

%\pacs{Valid PACS appear here}

\maketitle

\section{INTRODUCTION}\label{sec:intro}

One of the differences between laboratory and space laser
interferometer gravitational wave detectors is that, in the
laboratory, the two arms of the interferometer that is used to
detect changes in the spacetime geometry are maintained at
precisely equal lengths. Therefore, when the signals from the two
perpendicular arms are combined, the laser phase noise in the
differenced signals cancels exactly.  In space, a laser interferometer
gravitational wave detector such as LISA \cite{LISA} will have
free-flying spacecraft as the end masses, and precise equality of
the arms is not possible. Other methods must then be used to
eliminate laser phase noise from the system \cite{TintoArmstrong,AET}.
These methods involve a heterodyne
measurement for each separate arm of the interferometer and data
processing that combines data from both arms to generate a signal
that is free of laser phase noise. In a previous paper
(\cite{LHH}, hereafter called paper I), the sensitivity curves for
space detectors using these techniques were generated by
explicitly calculating transfer functions for signal and noise, as
modified by the data processing algorithms.  While the algorithms
have been shown \cite{AET}, in principle, to eliminate the laser
phase noise in the detectors regardless of the lengths of the two
arms, the transfer functions have previously only been calculated
for the case of equal arms \cite{LHH,Schilling,ETA,TEA}.  In this
paper we extend the calculation of the noise and signal transfer
functions to the case of arbitrarily chosen armlengths.

One of the goals of paper I was to provide a uniform system for
evaluating the sensitivity of various configurations of space
gravitational detectors.  This paper extends that capability to
configurations in which the armlengths are significantly different
from each other.  For example, a proposal by Bernard Schutz at the
2000 LISA Symposium in Golm, Germany \cite{Schutz}, suggested a
modification to the current LISA design in which a fourth spacecraft
is inserted in the middle of one of the legs of the interferometer to
produce two independent interferometers, each with one leg half the
length of the other (see Fig.~\ref{fig:Detector}).  The goal of such a
design was to be able to cross-correlate the independent
interferometers to search for the stochastic cosmic gravitational wave
background.  Using the analysis presented here, one will be
able to determine the sensitivity of such an interferometer and judge
the scientific value of the proposed modification.

As in paper I, the analysis begins with the response of a round-trip
electromagnetic tracking signal to the passage of a gravitational
wave, as derived by Estabrook and Wahlquist \cite{EstaWahl}.  A
gravitational wave of amplitude $h(t)$ will produce a Doppler
shift $\Delta\nu$ in the received frequency, relative to the
outgoing signal with fundamental frequency $\nu_0(t)$.  The shift
is given by
\begin{eqnarray}
      \frac{\Delta \nu (t,\theta ,\psi )}{\nu_{o}} &=& \frac{1}{2}
\cos
      2\psi  \nonumber \\
      &&\ \times \left[ (1-\cos \theta )h(t)+2\cos \theta ~h(t-\tau -\tau
      \cos \theta )-(1+\cos \theta )h(t-2\tau )\right] ,
      \label{DopplerSignal}
\end{eqnarray}
where $\tau $ is the one-way light travel time between spacecraft,
$\theta $ is the angle between the line connecting the spacecraft
and the line of sight to the source, and $\psi $ is a principal polarization angle
of the quadrupole gravitational wave.  It is desirable to work in
frequency space, so $h(t)$ is written in terms of its Fourier
transform $\tilde{h}(\omega)$.  If the Doppler record is sampled
for a time $T$ then $h(t)$ is related to its Fourier transform by
\begin{equation}
        h(t) = \frac{\sqrt{T}}{2 \pi} \int_{-\infty}^{+\infty}
        \tilde{h}(\omega) e^{i\omega t} d\omega \ ,
        \label{FourierTransform}
\end{equation}
where the $\sqrt{T}$ normalization factor is used to keep the power
spectrum roughly independent of time.  Using this definition of the
Fourier transform, the frequency shift of Eq.\ (\ref{DopplerSignal}) can
be written as
\begin{eqnarray}
        \ \Delta \nu (t,\theta ,\psi ) & = &
     \frac{\nu_{o}\sqrt{T}}{2 \pi}
        \int_{-\infty}^{+\infty}
        \frac{1}{2} \cos (2 \psi) \ \tilde{h}(\omega ,\theta ,\phi ,\psi)
        \nonumber \\
        & & \times \left[ (1 - \mu) + 2 \mu e^{-i \omega \tau(1 + \mu)} -
        (1 + \mu)e^{-i 2 \omega \tau} \right] e^{i \omega t} d\omega \ ,
        \label{DopplerSignal2}
\end{eqnarray}
where $\mu \equiv \cos \theta$.   The quantity that is actually read out
in a laser interferometer tracking system is phase, so
Eq.\ (\ref{DopplerSignal2}) is integrated to find the phase in cycles
\begin{equation}
      \Delta \phi (t,\theta ,\psi )=
      \displaystyle \int
      \Delta \nu (t,\theta ,\psi )\text{\ }dt.
      \label{PhaseSignal}
\end{equation}

In paper I, a strain-like variable $z$ was formed by dividing the
$\Delta\phi$ in Eq.\ (\ref{PhaseSignal}) by $\nu_0\tau$ and the
analysis was done using this variable.  Since both arms had
roughly the same length in paper I and carried nearly the same
frequency, there was only a scale difference between using
$\Delta\phi$ and using $z$ as the observable, and linear
combinations of $z$ were the same as linear combinations of
$\Delta\phi$.  However, when the two armlengths are different,
this is no longer the case, and one must be careful as to what is
taken to be the observable for use in noise-cancelling data
analysis.

In the laser phase-noise-cancellation algorithms that will be
presented in Section II, it is {\it relative phase} and {\it not strain}
that can be combined to create laser-noise-free signals. To understand how
this arises, consider a case where
laser signals in two arms are phase-locked to each other, with
$\nu_1$ as the frequency of the master laser in the first arm and
$\nu_2=\chi\nu_1$ as the frequency in the second arm, with
$\chi$ as the ratio of the two frequencies. Then a phase noise
excursion $\delta\phi_1$ in the first arm will produce a phase
noise excursion $\chi\delta\phi_1$ in the second arm. Thus it will
be linear combinations of $z_i=\phi_i/\nu_i$ that will allow the
two noise terms to cancel. Therefore, in this paper,
the gravitational wave observable in the $i^{\rm th}$ arm is defined
to be
\begin{eqnarray}
       z_i(t,\theta,\psi) & \equiv & \frac{\Delta\phi_i(t,\theta ,\psi )}
       {\nu_i}  \nonumber \\
       & = & \frac{\sqrt{T}}{4 \pi }
       \int_{-\infty}^{+\infty} d\omega \cos 2 \psi \ \tilde{h}(\omega)
       \left[ (1 - \mu) + 2 \mu
        e^{-i \omega \tau(1 + \mu)} - (1 + \mu)e^{-i 2 \omega \tau}
        \right] \frac{1}{\omega} e^{i \omega t} \ ,
        \label{zSignal}
\end{eqnarray}
where Eq.\ (\ref{DopplerSignal2}) has been used to expand $\Delta
\nu (t,\theta ,\psi )$ and where arbitrary constant phases have
been set to zero in the integration.  It should be noted that
$z_i$ is a different observable than the strain variable that was
labelled $z_i$ in paper I.  It should also be noted that $z_i$, as
it is now defined, has units of time, so Eq.\ (\ref{zSignal})
gives the {\it time delay} in seconds produced by the passage of a
gravitational wave through the detector.

\section{SENSITIVITY CURVES}\label{sec:SenseCurves}

\subsection{Instrument Signal}\label{sub:InstrumentSignal}
Tinto and Armstrong \cite{TintoArmstrong} originally showed that the
preferred signal for purposes of data analysis is not the traditional
Michelson combination (difference of both arms), but rather a new
combination $X(t)$, given in the time domain by \footnote{This form of
$X(t)$ assumes that the lasers in the end spacecraft are phase-locked
to the signals they receive from the central spacecraft. There is a
form for $X(t)$ that does not make this assumption and that can thereby
be converted to interferometers centered on the other spacecraft in the
constellation. However, the sensitivity we derive using this form will be
valid for the more general form as well, and will therefore apply to signals
formed with any spacecraft as vertex.}
\begin{eqnarray}
      X (t) &=& s_1(t) - s_2(t) - s_1(t-2\tau_{2}) + s_2(t-2\tau_{1})
      \nonumber \\
      &=& z_{1}(t) - z_{2}(t) - z_{1}(t - 2\tau_{2}) + z_{2}(t -
      2\tau_{1})  \nonumber \\
      &&\ \ \ + n_{1}(t) - n_{1}(t - 2\tau_{2}) - n_{2}(t) + n_{2}(t -
      2\tau_{1})\ ,
\label{XSignal}
\end{eqnarray}
where $s_{i}(t)$ is the data stream from the $i^{th}$ interferometer
arm, composed of the signal $z_{i}(t)$ of interest (given by Eq.\
(\ref{zSignal})) and the combined noise spectra in each of the
interferometer arms, $n_{i}(t)$.  The armlengths are taken to be
unequal, with armlength $\tau_{i}$ in the $i^{th}$ arm.  This
combination is devoid of laser phase noise for all values of the two
armlengths $\tau_{1}$ and $\tau_{2}$

To determine the sensitivity using the $X(t)$ variable, it is
necessary to establish a relationship between the amplitude of a
gravitational wave incident on the detector and the size of the $X(t)$
signal put out by the instrument.  The noise in the detector will limit
this sensitivity, and must also be included in the analysis.  The part
of $X(t)$ containing the gravitational wave signal
is\footnote{In paper I, the signal part of $X(t)$ was called $\Xi(t)$
in the limit where $\tau_{1} \rightarrow \tau_{2}$.}:
\begin{equation}
       \Lambda(t) = z_{1}(t) - z_{2}(t) - z_{1}(t - 2 \tau_{2}) +
       z_{2}(t - 2 \tau_{1})\ ,
          \label{LambdaDef}
\end{equation}
The transfer function $\cal R(\omega)$, which
connects the spectral density of the instrument output,
$S_{\bar{\Lambda}}(\omega)$ with the spectral density $S_{h}(\omega)$
in frequency space, is defined via
\begin{equation}
      S_{\bar{\Lambda}}(\omega) = S_{h}(\omega) {\cal R}
      \left(\omega\right) \ ,
      \label{TransferEquation}
\end{equation}
where the bar over the $\Lambda$ in Eq.\ (\ref{TransferEquation}) indicates
an average over source polarization and direction.
The gravitational wave amplitude spectral density $S_{h}(\omega)$
is defined by
\begin{equation}
          S_h(\omega) = | \tilde{h}(\omega) |^2 \ ,
\label{SpectralDefinition}
\end{equation}
where $\tilde{h}(\omega)$ is the Fourier amplitude defined in
Eq.\ (\ref{FourierTransform}), so that the mean-square gravitational
wave strain is given by
\begin{equation}
      \langle h^{2} \rangle = \frac{1}{T} \int_{0}^{\infty} h (t)^{2}
      dt = \frac{1}{2 \pi} \int_{0}^{\infty} S_{h}(\omega) d\omega \ .
      \label{SpectralDefinitionProps}
\end{equation}
Similarly, the instrumental response $S_{\bar{\Lambda}}(\omega)$ is
defined such that
\begin{equation}
      \overline{\langle \Lambda^{2} \rangle}
      = \frac{1}{2 \pi} \int_{0}^{\infty} S_{\bar{\Lambda}}(\omega)
      d\omega \ ,
      \label{SpectralPower}
\end{equation}
where the brackets indicate a time average.  In the next section, the
transfer function from the gravitational wave amplitude $h$ to the
instrument signal $\bar\Lambda$ is worked out.

\subsection{Gravitational Wave Transfer Function}\label{sub:Transfer}

Let us take the ratio of the two armlengths in the interferometer to be
an adjustable parameter, $\beta$, taking on values between $0$ and $1$,
such that $\tau_{1} = \tau$ and $\tau_{2} = \beta \tau$.  The average
power in the part of $X(t)$ which contains the gravitational wave signal
is given by
\begin{equation}
     \langle \Lambda^{2} \rangle = \lim_{T \rightarrow \infty} \frac{1}
     {T}
     \int_{0}^{\infty} \left| \Lambda \right|^{2} dt \ ,
     \label{PowerDef}
\end{equation}
where $\Lambda$ is defined by Eq.\ (\ref{LambdaDef}).  Using the
definition of $z$ from Eq.\ (\ref{zSignal}) this can be expanded
to yield
\begin{equation}
     \langle \Lambda^{2} \rangle = \frac{1}{2 \pi}
         \int_{0}^{\infty} d\omega \ \tilde{h}^{2}(\omega) \frac{1}
         {\omega^{2}} \left[ T_{1}(\omega) + T_{2}(\omega) - 2
         T_{3}(\omega) \right] \ ,
     \label{Power}
\end{equation}
where
\begin{eqnarray}
     T_{1}(u) & = & \cos^{2}(2 \psi_{1}) \cdot 4 \sin^{2}(\beta u)
     \left[ \mu_{1}^{2}\left(1 + \cos^{2}(u) - 2 \cos(u)\ \cos(u
     \mu_{1}) \right) \right.  \nonumber \\
     & - & \left. 2 \mu_{1} \sin(u)\ sin(u \mu_{1}) + \sin^{2}(u)
\right] \ ,
     \label{T1} \\
     T_{2}(u) & = & \cos^{2}(2 \psi_{2}) \cdot 4 \sin^{2}(u) \left[
     \mu_{2}^{2}\left(1 + \cos^{2}(\beta u) - 2 \cos(\beta u)\
     \cos(\beta u \mu_{2}) \right) \right.  \nonumber \\
     & - & \left.  2 \mu_{2} \sin(\beta u)\ sin(\beta u \mu_{2}) +
     \sin^{2}(\beta u) \right] \ , \label{T2} \\
     T_{3}(u) & = & \cos (2 \psi_{1}) \ \cos (2 \psi_{2}) \cdot 4
     \sin(u) \sin(\beta u) \ \eta(u)  \ ,
     \label{T3}
\end{eqnarray}
with $u = \omega \tau$, $\mu_{i} = \cos \theta_{i}$, and where
\begin{eqnarray}
      \eta(u, \theta_{1}, \theta_{2}) = \left[ \cos (u) - \cos
      (u \mu_{1}) \right] \left[ \cos (\beta u) - \cos
      (\beta u \mu_{2}) \right] \mu_{1} \mu_{2} \nonumber \\
      + \left[ \sin (u) - \mu_{1} \sin (u \mu_{1}) \right] \left[ \sin
      (\beta u) - \mu_{2} \sin (\beta u \mu_{2}) \right]
      \label{eta}
\end{eqnarray}
has been defined for convenience.  The propagation angles $\theta_{i}$
and principal polarization angles $\psi_{i}$ are defined with respect
to the $i^{th}$ arm using the geometric
conventions of paper I.  The expression for the power in the
detector, as given by Eq.\ (\ref{Power}), is a complicated function of
frequency and of the orientation between the propagation vector of the
gravitational wave and the interferometer, and represents the antenna
pattern for the detector.

It is customary to describe the average sensitivity of the instrument
by considering the isotropic power, obtained by averaging the antenna
pattern over all propagation vectors and all polarizations\footnote{The input
gravitational wave state given in Eq.\ (\ref{DopplerSignal}) is linearly
polarized. As in paper I, the averaging procedure over all linearly polarized
states produces the same response function as averaging over a more general
elliptically polarized state with an appropriately weighted distribution}.  Using the
definition of $\cal R(\omega)$ from Eq.\ (\ref{TransferEquation}), with
the average isotropic power computed using the geometric averaging
procedure of paper I with Eqs.\ (\ref{T1} - \ref{eta}), the
gravitational wave transfer function is found to be
\begin{eqnarray}
      {\cal R}(u) & = & \left(\frac{\tau}{u}\right)^{2} \left\{ 2
      \sin^{2}(\beta u) \left[ \left(1 + \cos^{2} (u) \right) \left(\frac{1}
      {3} - \frac{2}{u^{2}} \right) + \sin^{2} (u) + \frac{4}
      {u^{3}} \sin (u) \ \cos (u) \right] \right.  \nonumber \\
          & + &  \left. 2 \sin^{2}(u) \left[ \left(1 + \cos^{2} (\beta u)
          \right) \left(\frac{1}{3} - \frac{2}{(\beta u)^{2}} \right) +
          \sin^{2} (\beta u) + \frac{4}{(\beta u)^{3}} \sin (\beta u) \
          \cos (\beta u) \right] \right. \nonumber \\
          & - & \left.  \frac{1}{\pi} \sin(u) \sin(\beta u)
          \int_{0}^{2\pi} d\epsilon \int_{-1}^{+1} d\mu_{1} \ \left(1 - 2
          \sin^{2} \alpha \right) \eta(u, \theta_{1}, \theta_{2})
\right\}
\ .
\label{TransferFunction}
\end{eqnarray}
The remaining integral can be evaluated using simple numerical
techniques, after relating the angular variables as described in
Paper I, where:
\begin{equation}
      \sin \alpha = \frac{\sin \gamma \ \sin \epsilon}{\sqrt{1 -
      \mu_{2}^{2} }} \ ,
      \label{AlphaDefn}
\end{equation}
and
\begin{equation}
       \mu_{2} = \mu_{1}\ \cos \gamma + \sin \gamma \ \cos \epsilon \
       \sqrt{1 - \mu_{1}^{2}} \ .
          \label{Mu2Defn}
\end{equation}
Here $\gamma$ is the opening angle of the interferometer, and $\epsilon$
is the inclination of the gravitational wave propagation vector to the
plane of the interferometer.
The complete gravitational wave transfer function is plotted in
Fig.~\ref{fig:GWTransfers} for $\beta = 1$ (``equal arm'') and
Fig.~\ref{fig:GWTransfers2} for $\beta = 0.01$ (``unequal arm'') examples.

As may be seen in the figure, the low-frequency (small $u$) response
of the detector to a gravitational wave signal is four orders of
magnitude lower for the $\beta=0.01$ detector than for the equal arm
detector, implying that the (amplitude) signal will be two orders of
magnitude lower -- the detected signal level is proportional to the
length of the shortest arm.  However, once the period of the
gravitational wave falls inside the light-time of the longest arm, $u
\sim 1$, the equal-arm detector ($\beta = 1$) response begins to fall
off while the unequal-arm detector ($\beta = 0.01$) response is
roughly flat up to a period corresponding to the light-time in the
shortest arm.

The dropoff at low frequencies is a result of the fact that the
variable $X(t)$ is formed by subtracting each $z_i$ from itself,
offset by the light-time in the opposite arm.  Thus, in the
low-frequency limit, the two copies of the signal strongly overlap and
the signal is almost entirely subtracted away.  For equal arms, the
response of the detector is likewise subtracted to zero when an
integer number of wavelengths fits in the arm length, as seen in the
high-frequency portion of the $\beta=1$ curve.  For the unequal-arm
case, this does not occur, because the subtraction of two versions of
the signal in each arm are done at different light-times in the two
arms, so whatever period signal cancels in one arm will not cancel in
the other.  However, as may be seen in the $\beta = 0.01$ case, the
response drops sharply to zero at $\log u \simeq 2.5$ (equivalent to
$f \sim 10^{0.5}$ Hz for LISA armlength of $c \tau = 5 \times 10^{9}$
m), where exactly one wavelength fits into the short arm and exactly
one hundred fit into the long arm.

However, the response of the detector's $X(t)$ signal is not the
whole story.  The ability of a detector to detect a signal depends
on both the signal in the detector and on the competing noise.
As we shall see in the next section, when the $X(t)$ variable is
formed, the noise in each arm is likewise subtracted away in most
of the places where the signal is lost ({\it e.g.}, at low frequency),
so the ratio of signal to noise remains high.

\subsection{Noise transfer function using the X(t) variable}
\label{sub:NoiseTransfer}

The noise sources for LISA may be divided into categories in two
different ways. First, a noise source may be either one-way (affecting only the
incoming or the outgoing signal at a spacecraft, but not both) or two-way,
(affecting both incoming and outgoing signals at the same time).  A one-way
 noise source will
have a transfer function of $2$, since there are $2$ spacecraft in each leg
contributing equal amounts of such noise \footnote{This choice of putting the
factor of 2 into the transfer function differs from our convention in paper I,
where such factors were included in the noise spectra. We have found it clearer
to define the transfer function as the one that gives the noise power spectrum
in
$X$ as $S_X = R S_N$, where $S_N$ is the noise spectral density of a single type
in a single spacecraft}. The transfer function for two-way noise sources,
however,
will be more complicated due to the internal correlation.  A single two-way
noise fluctuation in the central spacecraft of the interferometer will
affect the incoming signal immediately, and then, a round-trip light-time
later, will affect the measured signal again in the same way. In
the time domain, the effect in the $i^{th}$ arm of a fluctuation $n(t)$
will be $n_i(t)=n(t)+n(t-2\tau_i)$.  The transfer function for this
time-delayed sum is $4\cos^2(2\pi f\tau_i)$.  If an end spacecraft has
noise that affects both incoming and outgoing beams, it will
affect them at almost the same time, with no delay, giving a transfer
function contribution of 4.  The noise transfer function for a single arm
for a two-way noise source is therefore
\begin{equation}
    4+4\cos^2(2\pi f\tau_i) \ .
   \label{TwoWay}
\end{equation}
Examples of one-way noise are thermal noise in the laser receiver
electronics  or a mechanical change in the optical pathlength in the
outgoing laser signal  before it gets to the main telescope optics.
Examples of two-way noise are parasitic forces on the accelerometer
proof mass or thermal changes in the optical pathlength in the main
telescope.

A second way in which noise sources may be classified is by how they
scale when there is a change in armlength in the interferometer.  The
first type of noise in this classification scheme is what we call
``position noise'', in which the size of the noise in radians of phase is
independent of the length of the arm. Accelerometer noise and thermal noise in
the laser electronics are examples of position noise.  The second type of noise
is what we call ``strain noise'', in which the size of the noise scales with
armlength.  Examples of strain noise include shot noise and pointing jitter (if
it is dominated by low power in the incoming beacon).  Position noises may be
either one-way or two-way, but we can think of no two-way strain noise sources.

The transfer functions that connect the noise in the instrument to the
$X$ variable depend on the type of noise.  We begin by considering the
noise terms in Eq.\ (\ref{XSignal}):
\begin{equation}
     \sigma(t) = n_{1}(t) - n_{2}(t) - \left[ n_{1}(t - 2 \beta \tau) -
     n_{2}(t - 2 \tau) \right] \ .
     \label{NoisePart}
\end{equation}
We then go to the frequency domain, squaring and time-averaging to obtain
the power spectrum.
\begin{equation}
     \langle \sigma^{2} \rangle = {1 \over {2 \pi}} \int d\omega \ 4
     \left[ \tilde{n}_{1}^{2} \sin^{2}(\beta u) + \tilde{n}_{2}^{2}
     \sin^{2}(u) \right] \ ,
     \label{NoisePower}
\end{equation}
where cross-terms ({\it e.g.}, $\tilde{n}_{1}\tilde{n}_{2}$) have
been neglected under the assumption that noise in the two arms will be
independent and uncorrelated.  Note that $\tilde{n}_{1}^{2}$ is the
power spectrum in the long arm (length $\tau$) and $\tilde{n}_{2}^{2}$
is the power spectrum in the short arm (length $\beta \tau$).

Since the noise in the detectors includes different types, with
different transfer functions, it is not possible to write a single transfer
function giving the response of the $X$ variable to noise, so let us consider
the various noise categories one at a time.  We first consider position
noise, for which $\tilde{n}^{2} \equiv \tilde{n}_{1}^{2} = \tilde{n}_{2}^{2}$.
Then, using Eq. (\ref{NoisePower}), we find the
transfer function for one-way position noise to be
\begin{equation}
     {\cal R}_{1} = 8 \left( \sin^{2}(\beta u) + \sin^{2}(u) \right)\ ,
     \label{PositionTransfer}
\end{equation}
where, as we noted above, there is a factor of $2$ representing the noise
from the two spacecraft in each arm.
Two-way position noise must include the transfer function from Eq.
(\ref{TwoWay}), giving
\begin{equation}
     {\cal R}_{2} =
           16 \left[ \sin^{2}(\beta u)\left(1+\cos^2(u)\right)
                + \sin^{2}(u)\left(1+\cos^2(\beta u) \right)\right]\ .
     \label{TwowayTransfer}
\end{equation}
Strain noise scales with armlength, and is hence smaller in the shorter arm,
so that $\tilde{n}^{2} \equiv \tilde{n}_{1}^{2} = \tilde{n}_{2}^{2}/\beta^2$.
Its transfer function is therefore
\begin{equation}
     {\cal R}_{s} = 8 \left( \sin^{2}(\beta u) + \beta^{2}
       \sin^{2}(u) \right)\ ,
     \label{ShotTransfer}
\end{equation}
where the factor of $2$ for the two spacecraft has again been included.

When $\beta=1$, the transfer functions for strain noise and one-way
position noise (Eqs. \ref{PositionTransfer} and \ref{ShotTransfer}) are
identical and have zeros at $u_n=n\pi$, where $n$ is zero
or a positive integer.  These are exactly the places where the
$\beta=1$ transfer function for gravitational wave signal (Fig.
~\ref{fig:GWTransfers}) has its zeros.  When $\beta < 1$, the situation is
more complicated.  Both ${\cal R}_{1}$ and ${\cal R}_{s}$ share the
$\sin^2(\beta u)$ term which will go to zero at $u=0$ and at multiples
of $u=\pi/\beta$.  The $\sin^2(u)$ terms in ${\cal R}_{1}$ and
${\cal R}_{s}$ have their zeros at multiples of the lower frequency,
$u=\pi$.  In ${\cal R}_{1}$, this term will be larger than
$\sin^2(\beta u)$ term at low frequencies, since near $u=0$,
$\sin^2(u)\simeq u^2$, while $\sin^2(\beta u) \simeq
\beta^2 u^2$. In ${\cal R}_{s}$, these terms will be
equal in the low-frequency limit, because of the factor $\beta^2$ that
multiplies the $\sin^2(u)$ term. Thus, in the low frequency limit,
the strain noise transfer function will be $2\beta^2$ times the
the one-way position noise transfer function. When $\beta \ll 1$,
the transfer function for one-way position noise will have sharp
drops at multiples of $u=\pi$, down to the level of its
$\sin^2(\beta u)$ term. These behaviors are shown in
Fig.~\ref{fig:NoiseTransfer} and Fig.~\ref{fig:NoiseTransfer2}.

\subsection{Sensitivity curve}\label{sub:sensitivity}

The signal to noise ratio is the ratio of the signal power in
the detector to the noise power in the detector:
\begin{equation}
     {\rm SNR} = \frac{S_{h} {\cal R}}{S_s{\cal R}_s
     +S_1{\cal R}_1+S_2{\cal R}_2}\ .
     \label{SNR}
\end{equation}
where $S_s$, $S_1$, and $S_2$ are the spectra of strain noise and
one-way and two-way position noise, respectively, and ${\cal R}$ is
the gravitational wave transfer function given by Eq.\
(\ref{TransferFunction}).
Setting ${\rm SNR} = 1$ and solving for $h_{f}\equiv\sqrt{S_{h}}$
yields the instrument sensitivity curve as defined in paper I:
\begin{equation}
      h_{f} = \sqrt{S_{h}} = \sqrt\frac{S_s{\cal R}_s +
      S_1{\cal R}_1 + S_2{\cal R}_2}
      {\cal R } \ .  \label{hf}
\end{equation}
where ${\cal R}$ is the gravitational wave transfer function, given by
Eq. (\ref{TransferFunction}).

Figures \ref{fig:EqualArmSense} and \ref{fig:UnequalArmSense} show the
sensitivity curves, computed using Eq.\ (\ref{hf}), for
$\beta = 1$ and $\beta = 0.01$ respectively.  The noise
values used are taken to be the LISA target design values
(computed as described in paper I).  The shot noise and
acceleration noise levels are set at the standard LISA values.
In addition, a flat one-way position noise spectrum is assumed at
$1/10^{th}$ the LISA shot-noise value.  Also plotted in Figures
\ref{fig:EqualArmSense} and \ref{fig:UnequalArmSense} are sensitivity curves
representing each of the three components of the total noise, taken
one at a time.

\section{DISCUSSION}\label{sec:discussion}

As may be seen in Fig. \ref{fig:UnequalArmSense}, the low-frequency
sensitivity for unequal
arms, being set by the two-way position noise in the accelerometer, is
degraded over the equal-arm case by the ratio of the two arms.  In other
words, the sensitivity at lowest frequencies is set by the sensitivity
of the shortest arm.  At middle and high frequencies, the situation is
more complicated.  If the dominant noise is strain noise, then the
sensitivity is independent of $\beta$ in this frequency range.
However, if the dominant noise is position noise,
then the sensitivity curve at high frequencies will rise in proportion
to $\beta$, though its flat floor will extend to higher frequency, from
the $1/(2\pi\tau)$ of the equal-arm case to $1/(2\pi\beta\tau)$ when the
armlength ratio is $\beta$.

The implications of these results for mission design are obvious.
If the armlengths are not equal, the low-frequency sensitivity is
degraded by a factor $1/\beta$, the ratio of the armlengths.  If
the high-frequency noise can be guaranteed to be strain noise, even
in the shorter arm, then the high-frequency sensitivity is unaffected
by the unequal arms.  If the noise at high frequency is dominated by
position noise, then the high frequency sensitivity is degraded by the
factor $1/\beta$, but the sensitivity remains flat up to a frequency
$1/(2\pi \beta\tau)$, where it turns over and joins the strain noise
curve.  Thus, as long as the position-noise sources can be kept well
below the shot noise and other strain-noise contributions, a change in
armlength ratio from strict equality will not degrade the high-frequency
portion of the sensitivity curves.  However, as the length of one of the
arms is shortened, small position noise sources will become important
and eventually dominate.

Let us consider the example of Schutz's 4-spacecraft configuration
(Fig.~\ref{fig:Detector}).  Since this configuration will have $\beta=0.5$,
the low-frequency sensitivity curve will be a factor of 2 higher
(hence less sensitive).
The current error budget for LISA assumes that the high-frequency
portion of the window is dominated by position noise approximately
three times the shot noise.  If this remains the case, then the
high-frequency section of the curve will likewise be a factor of 2
higher up to a frequency twice as high as the LISA sensitivity
``knee'' at $f = 1/(2\pi \tau)$,
at which point it would turn up and join the current LISA
high-frequency ramp.  The shot noise is determined by the power of
the laser and by the size and efficiency of the optics, and there
is nothing beyond brute-force improvements in these parameters
that will lower the shot noise.  The contributions to position
noise, on the other hand, are due to optics quality, the attitude
control system, Brownian noise in the electronics, thermal noise
in the optical path length, {\it etc}.  These are more complex
and are amenable to reduction by careful or innovative engineering
design.  If these noise sources can be reduced to a fraction of
the shot noise, not only will the LISA noise floor be reduced by a
factor of 4, but the Schutz modification will have high-frequency
performance that is undiminished by the reduction of the length of
one arm.

Finally, we describe a totally unfeasible mission design that is
nevertheless interesting for instructive purposes.  Let us
consider a two-spacecraft ``interferometer'', where one of the
spacecraft contains a fiber optic delay line, of length 5 km, that
acts as the second arm of the interferometer.  If the distance
between the two spacecraft is $5 \times 10^6$ km, we will have
$ \beta = 10^{-6}$.  The use of the $X(t)$ variable will eliminate
laser phase noise, exactly as it does in arms that are more nearly
equal.  A rigidly-attached reflector at the far end of the
fiber-optic line would eliminate accelerometer noise, but, of course,
would replace it with thermal fluctuation in the optical path
length in the fiber. However, a concatenation of fibers with
well-chosen thermal pathlength coefficients could produce a fiber
tuned to have a coefficient very near zero.  This, combined with
multilevel thermal isolation, could keep this noise source very
small.  The key to the sensitivity of this configuration is the
position noise.  If a way can be found to reduce position noise
to less than $10^{-6}$ of the LISA shot noise, then this
two-spacecraft interferometer would have the same sensitivity as
a conventional three-spacecraft interferometer.

\begin{acknowledgments}
S.\ L.\ L.\ acknowledges support for this work under LISA contract
number PO 1217163, and the NASA EPSCoR Program through
Cooperative Agreement NCC5-410. The work of W.\ A.\ H.\ was
supported in part by NSF Grant No. PHY-0098787 and the
NASA EPSCoR Program through Cooperative Agreement NCC5-579.
R.\ W.\ H.\ was supported by NASA grant NAGS5-11469 and NCC5-579.

\end{acknowledgments}
% ==========================================================

% ===== FIGURES ======================================
\newpage

\begin{figure}
      \caption{\label{fig:Detector} An unequal arm geometry used here assumes two arms of
      length $\tau$ and $\beta \tau$, with an enclosed angle
     $\gamma$ (the
      interferometer opening angle).  Depicted here is the nominal LISA
      constellation of three spacecraft in an equilateral triangle, and a
      proposed extension which places a fourth spacecraft midway down one
      of the arms.}
\end{figure}

\begin{figure}
      \caption{\label{fig:GWTransfers} The dimensionless gravitational wave transfer
      function, ${\cal R}/\tau^2$, plotted against the dimensionless frequency parameter
      $u = \omega \tau$, for value of $\beta = 1.0$.}
\end{figure}

\begin{figure}
      \caption{\label{fig:GWTransfers2} The dimensionless gravitational wave transfer
      function, ${\cal R}/\tau^2$ plotted against the dimensionless parameter
      $u = \omega \tau$, for value of $\beta = 0.01$.}
\end{figure}

\begin{figure}
      \caption{\label{fig:NoiseTransfer} The noise transfer functions for $\beta = 1$
      as functions of the dimensionless frequency parameter $u = \omega \tau$.
      Notice that the transfer function for position noise (${\cal R}_{1}$)
      is identical to the transfer function for shot noise (${\cal R}_{s}$)in
      the $\beta = 1$ limit.}
\end{figure}

\begin{figure}
      \caption{\label{fig:NoiseTransfer2} The noise transfer functions for
      $\beta = 0.01$ as functions of the dimensionless frequency parameter
      $u = \omega \tau$.}
\end{figure}

\begin{figure}
      \caption{\label{fig:EqualArmSense} The sensitivity curve ($SNR = 1$) for $\beta = 1$.
      Overlayed are the sensitivity curves for each of the individual
      noise spectra (acceleration noise, shot noise, position noise).
      The noise spectra are taken to be at the LISA target design values,
      except position noise, which is taken to be $1/10^{th}$ the LISA
      value.}
\end{figure}

\begin{figure}
      \caption{\label{fig:UnequalArmSense} The sensitivity curve ($SNR = 1$) for $\beta = 0.01$.
      Overlayed are the sensitivity curves for each of the individual
      noise spectra, as in the previous figure.}
\end{figure}

\end{document}